\documentstyle[aps,prb,twocolumn,epsfig]{revtex}  
\begin{document}
\draft

\twocolumn[\hsize\textwidth\columnwidth\hsize\csname @twocolumnfalse\endcsname

\title{Possible Exotic phases in the One--Dimensional 
Extended Hubbard Model}

\author{R.~Torsten Clay, Anders W.~Sandvik, and David K.~Campbell}
\address{Department of Physics, University of Illinois 
at Urbana-Champaign, 1110 W.~Green Street, Urbana, Illinois 61801}

\date{\today}

\maketitle

\begin{abstract}
We investigate numerically the ground state phase diagram of the 
one-dimensional extended Hubbard model, including an on--site
interaction $U$ and a nearest--neighbor interaction $V$.  
We focus on the ground state phases of the model in the $V \gg U$
region, where previous studies have suggested the possibility of dominant 
superconducting pairing fluctuations before the system phase 
separates at a critical value $V=V_{\rm PS}$. 
Using quantum Monte Carlo methods on lattices much
larger than in previous Lancz\"os diagonalization studies,
we determine the boundary of phase separation, the Luttinger Liquid 
correlation exponent $K_\rho$, and other correlation functions in
this region. We find
that phase separation occurs for $V$ significantly smaller than 
previously reported. In addition, for negative $U$, we find that a
uniform state re-enters from phase separation as the electron
density is increased towards half filling. For $V < V_{\rm PS}$, our 
results show that superconducting fluctuations are not dominant. The
system behaves asymptotically as a Luttinger Liquid with $K_\rho < 1$, 
but we also find strong low-energy (but gapped) charge-density
fluctuations at a momentum not expected for a standard Luttinger Liquid.
\end{abstract}

\pacs{PACS numbers: 71.10, 74.20.Mn}]

\section{Introduction}
One-dimensional (1D) Hubbard models have been used to model many 
quasi-1D systems, including conducting polymers such as polyacetylene,
\cite{condpoly} and organic charge--transfer materials such as TTF-TCNQ or 
${\rm (TMTSF)_2 PF_6}$.\cite{orgsc} In the simplest form, with only an 
on--site interaction $U$, the 1D Hubbard model has been solved exactly 
using the Bethe {\it Ansatz}.\cite{lw} However, longer range interactions 
are needed to describe many effects observed in real materials, e.g., the 
excitons found in conducting polymers.\cite{exciton1} The conventional 
extended Hubbard model (EHM), which in addition to 
 $U$ includes a nearest-neighbor interaction of strength $V$,
is the simplest extension that takes into account some of the longer-range
interaction effects.
The EHM Hamiltonian is
\begin{eqnarray}
H&=&-t \sum_{i \sigma} (c_{i+1 \sigma}^\dagger c_{i\sigma} + 
        c_{i\sigma}^\dagger c_{i+1\sigma}) 
+ U \sum_{i} n_{i\uparrow}n_{i\downarrow} \nonumber \\
& & +V \sum_{i} n_i n_{i+1},
\end{eqnarray}
where $c^\dagger_{i\sigma}$ creates an electron of spin $\sigma$ on site $i$, 
and $n_{i\sigma}=c^\dagger_{i\sigma}c_{i\sigma}$ ($n_i=n_{i\uparrow}+
n_{i\downarrow}$) are electron number operators. An implicit
parameter is the filling factor $n= N_e/N$, where $N_e$ and $N$ are 
respectively the number of electrons and lattice sites. We will henceforth
give energies in units of the hopping $t$.

As one of the basic many-body Hamiltonians in 1D, the EHM has been the subject 
of a large number of studies.\cite{hs,mz,pm,so,big2,big} Nonetheless,
as we shall discuss below, there remain important open questions
related to the phase diagram at intermediate and strong coupling, where
both analytical and numerical methods are difficult to apply reliably.
One of the principal reasons for the existence of these open questions
is the variety of potential broken symmetry fluctuations that
can occur in the EHM. As the parameters $U$, $V$, and $n$ are varied, 
several different types of correlations dominate the ground state, including 
spin or charge density wave (SDW/CDW), and singlet or triplet superconducting
(SC) fluctuations. Of course, in a strictly 1D model, long-range order that
breaks a continuous symmetry (i.e., SC or SDW)
is not possible; however long-range CDW order can occur at zero temperature. 
Further, in some parameter regions, the EHM is unstable towards 
phase separation (PS), with the nature of the coexisting phases depending
on the parameters.

For small values of the interaction parameters, the low-energy
excitations of the
EHM can be mapped directly to a 
``Luttinger liquid'' (LL), \cite{ll1} the unifying framework for 
1D interacting fermion systems obtained from weak-coupling 
renormalization group studies, bosonization, and conformal field theory.
The general ($q$-component) LL contains $q$ gapless degrees of freedom,
and the forms of the correlation functions depend on only two parameters 
for each gapless mode $\alpha$: A renormalized velocity $v_\alpha$
and an interaction parameter $K_\alpha$. For instance, the standard 
Tomonaga-Luttinger model \cite{luttinger,tomonaga} has
two gapless degrees of freedom, charge and spin, and is thus a
two-component LL with interaction parameters $K_{\rho}$ and
$K_{\sigma}$; a similar one--dimensional exactly solvable model, due to 
Luther and Emery, \cite{le} has gapless charge excitations, but a spin gap, 
and thus behaves as a Luttinger liquid
only in the charge sector. The identification of a 
given model as a Luttinger liquid
enables (in principle) a straightforward numerical 
investigation of the ground state phases of the model. In the case of the 
integrable standard Hubbard model,
the Bethe {\it Ansatz} equations have been used this way to calculate
the ground state parameters and $K_\rho$ for all values of the repulsion $U$
and the band fillings.\cite{schulz} For more general models, away from weak 
coupling, there is unfortunately no reliable analytic method to determine the
parameters, although they may in principle be calculated numerically.
Assuming the model remains a LL also away from the weak-coupling regime, 
numerical methods can be used to calculate the exponents $K_\alpha$, 
which will then give the asymptotic form of the correlation functions. 
Importantly, for such calculations to be meaningful, finite-size and other 
systematic errors must be carefully analyzed, and in particular, one must 
ascertain that the model is {\it not} phase separated in the thermodynamic 
limit. As we demonstrate below, calculating Luttinger liquid
parameters from finite-size
systems can sometimes be highly problematic, particularly close to
phase separation.

The existence of several distinct regions and types of phase separation
in the phase
diagram of the EHM \cite{mz,pm,big2,big} and
other 1D strongly-correlated models \cite{pssc1,pssc2,cuo1,cuo2} is
well-established. While phase separation typically occurs outside the 
parameter regions thought to be relevant for modeling real materials,
for purposes of understanding the behavior of any given theory it is imperative 
that all phase-separated regions be identified and, if possible,
understood prior to 
carrying out other studies, such as determinations of Luttinger liquid
correlation 
exponents. In small systems, the signals for phase separation
can be unclear or ambiguous. 
For instance, the correlation functions calculated for small systems can be 
misleading in phase-separated regions,
since the boundary between the two phases can 
be large compared to the system size. This leads to a mixing of correlations 
from the two different phases, which may have quite different properties 
(for example, spin-gapped and non-spin-gapped). Nonetheless, in a
small system, these correlations may not show signs of phase separation
until well inside the phase-separated region. Sometimes, it may then
appear that the
exponent $K_\rho>1$ (which would signal dominant superconducting
correlations), when 
in fact the system is phase separated in the thermodynamic limit and
is thus not a uniform Luttinger liquid. Accordingly, one of our principal
goals in this 
study has been to understand thoroughly the phase-separated regions before
calculating Luttinger liquid exponents or other correlation functions.

Recently, the extended Hubbard model
in the region of parameters $V \gg U \sim t$ has 
attracted considerable interest because of the possibility of a novel
superconducting ground state away from half-filing ($n < 1$).\cite{mz,pm,so}
Although the interactions appear to be purely 
repulsive when $V>U>0$ (we also consider $V\gg U$ with $U<0$), the gradient 
of the potential is positive, and there is an attractive {\it force} between 
electrons. Pairs might then form and could in principle lead to a ground state 
with dominant superconducting
fluctuations. Based on Lancz\"os exact diagonalization (ED)
results at quarter--filling ($n=1/2$), it was argued that superconducting
fluctuations 
indeed dominate for a substantial range of $V$ values, before the system 
phase separates at a very large $V=V_{\rm PS}$.\cite{mz,pm}
Another ED study found similar behavior at filling $n=2/3$, but no
attempt was made to determine whether the system is phase separated
at this filling.\cite{so} These previous studies also addressed the 
question of the location of the spin gap boundary. Determining the location
of the spin-gapped Luther-Emery phase is important 
to studying the possibility of superconductivity
in this region, as the presence of a spin 
gap would be consistent with the expected dominant singlet pairing 
correlations when $V$ is large and positive. Superconducting
correlations involving 
triplet pairs have also been proposed as a scenario to explain ED
results.\cite{pm} Triplet superconducting
correlations are dominant in some regions with 
$V<0$ \cite{big2,big}), but explaining their origin for $V \gg |U|$ is
problematic.

In this paper, we investigate the related questions of novel superconducting 
fluctuations, calculations of $K_{\rho}$, and the boundaries of phase
separation and 
spin-gapped regions using quantum Monte Carlo (QMC) simulations of relatively
large systems (up to 128 sites). We focus on the ill-understood
parameter region $V \gg |U| $ for a wide range of fillings. Our results
show that phase separation extends to much lower values of $V$ than previously 
reported.\cite{mz,pm}  We also find that for negative $U$ the
high-density phase is not the naively expected one with doubly occupied 
sites separated by one site (corresponding to half filling in the
high-density phase) but is at a lower density. A uniform
state (which has strong CDW fluctuations) re-stabilizes as half filling 
is approached. In most of the parameter space we can conclude 
that $K_\rho<1$ for $V < V_{\rm PS}$. In some cases, our results are on the 
border--line $K_\rho \agt 1$ at our calculated phase separation boundary,
but in no case do we find a definite $K_\rho > 1$ before phase separation.
We therefore believe that
superconducting correlations never dominate in the $V \gg |U|$ 
region. Instead, for a range of parameters, we find strong charge-density 
fluctuations at a momentum $2k_{\rm F} < q < 4k_{\rm F}$, which are not 
expected in a standard Luttinger liquid.
Our analysis of the temperature and size dependence of these
fluctuations shows
that they do not correspond to gapless excitations, and hence
the model remains a Luttinger liquid
in the asymptotic low-energy sector. Nevertheless, 
these strong anomalous fluctuations demonstrate the appearance of non-Luttinger
liquid features in the excitation spectrum.

To present our results, we begin in  Sec.~II by discussing briefly 
the numerical methods we have used to study the extended Hubbard
model. We point out advantages
and disadvantages of several different QMC methods and stress the
necessity of comparing and contrasting their predictions to
obtain definitive conclusions. In Sec. III, we discuss 
methods to detect phase separation in numerical data, emphasizing a number
of often overlooked subtleties, and present our results
for the phase separation boundaries of the extended Hubbard
model. We address the determination of Luttinger liquid exponents
and other correlation functions in Sec. IV. Sec.~V concludes with a summary 
of our understanding of the ``exotic'' phases of the extended
Hubbard model in the region $V\gg |U|$.

\section{Numerical Methods}

Using the Luttinger liquid formalism, extracting the dominant
correlations of 1D electron
systems is easy in principle. Lancz\"os exact diagonalization of small
systems can give results for certain observables (e.g., velocities and
stiffness constants) that can be used to determine the asymptotic
form of the correlations functions which are believed to be
less affected
by finite-size effects than the correlation functions themselves. This has 
led to an upsurge of ED studies of various 1D models, including many focusing
on possibilities of dominant superconducting fluctuations close to phase 
separation.\cite{mz,pm,so,pssc1,pssc2,sudbo,dagotto,ekl}
However, as we argue below, the finite-size effects may in fact be unexpectedly
large in regions near phase separation. It is therefore important to
confirm ED results using methods that can treat larger system sizes. For 
this purpose, we have used three different QMC methods. Here we 
summarize briefly the basic ideas of these techniques, referring readers to the 
literature for additional details. Our main purpose is to make some observations 
 concerning the advantages and disadvantages of the
different QMC methods in studies of the EHM in the difficult parameter regime
$V \gg |U|$. We believe that most of this discussion will apply to the 
region of very strong interactions in other models as well.

The first method is based on the ``stochastic series expansion'' (SSE) of the 
density matrix $e^{-\beta H}$.\cite{sse1,sse2} This is a generalization of 
Handscomb's method,\cite{handscomb} applicable for a much larger class 
of lattice Hamiltonians. There are no approximations such as Trotter 
discretization of imaginary time, but in order to obtain ground state results
one has to ensure that a large enough inverse temperature $\beta$ is used. 
To check for finite-temperature effects, we have carried out 
calculations for several values of $\beta$. In general, we find that 
$\beta \sim 2N-4N$ is sufficient to give accurate values for the
ground state parameters for $N$ up to $128$.

The second technique is a recently developed variant of the continuous-time
worldline algorithm proposed by Prokofev {\it et al.}.\cite{prokofev}
Our version of this method \cite{irsse} uses an updating scheme adapted from 
the SSE algorithm (the two methods are, in fact, closely related to each 
other \cite{irsse}). The algorithm is based on the finite-temperature 
perturbation expansion in the interaction representation, around the 
atomic-limit system with no kinetic energy term. This expansion converges 
for a finite system at finite $\beta$. The terms (which are paths in 
continuous imaginary time) can be stochastically sampled in much the
same way as is done in the SSE method. We will refer to this QMC 
technique as the interaction representation (IR) method. It is 
also free from systematic errors.

Results of SSE and IR simulations are in general in complete agreement with
each other. Both methods operate in the real-space occupation number
basis and hence suffer from well known ``sticking problems'' (inability
of the local Monte Carlo updates to evolve the real space configurations
through states with high potential energy) when the interactions are very 
strong. For the EHM, $V \approx 10$ seems to be the highest accessible $V$
in the interesting 
filling regions ($U$ represents a lesser problem, since we consider 
$V \gg U$). We have found that the autocorrelation times for spin and
density correlation functions at strong interactions are shorter for the IR 
method, in particular close to half filling. All results presented here 
for correlation functions and susceptibilities
were therefore obtained with that method. 
However, for the ground state energy (i.e., the internal energy at 
sufficiently high $\beta$), the SSE method gives results with statistical 
errors approximately an order of magnitude smaller than the IR method (in 
cases where the sticking problems do not make simulations practically 
impossible). This result probably arises because the total energy estimator 
of the SSE method is directly related to the stochastically averaged order 
of the terms of the expansion of $e^{-\beta H}$. In the IR method (and other 
worldline methods), the energy estimator consists of separate diagonal and  
off-diagonal parts, i.e., the potential and kinetic energies are not 
treated on an equal footing (in particular, the diagonal part is typically
not spin-rotationally invariant). We note that the SSE method also has 
proven superior in energy calculations for spin models, such as the 
two-dimensional Heisenberg model,\cite{hberg2d} for which otherwise 
very efficient loop algorithms \cite{loops} have not given nearly as 
accurate results.

The third QMC method we employ is the recently developed Constrained 
Path Monte Carlo (CPMC) algorithm. This is a Slater-determinant based projector
method that handles the interactions via a Trotter decomposition and a
Hubbard-Stratonovich transformation, leading to auxiliary fields that are 
sampled.\cite{cpmc1,cpmc2} A constraining wavefunction is used to prevent 
the fermion sign problem. In 1D, the constraint becomes exact
(i.e., there are no sign problems). There is a small 
systematic error originating from the Trotter decomposition, which can be 
made arbitrarily small by decreasing the ``imaginary time slice'' width. 
In our comparisons with SSE and IR results, CPMC yielded similar results, 
except at very strong interactions where matrix conditioning problems
become overwhelming and make the method very hard to use (much before the
sticking problems become problematic in SSE and IR simulations). 
One advantage of CPMC over the SSE and IR methods is that 
in CPMC any equal-time correlation function
may be computed since the Green's function is directly
accessable in CPMC.
To obtain  accurate energies from CPMC it was necessary to
compute the energy for more than one $\Delta\tau$ value and scale
the results to $\Delta\tau\rightarrow 0$. Our implementation of
CPMC used a uniform free-electron wavefunction for the
constraint, the initial wavefunction, and the importance sampling
wavefunction. While the choice of the importance sampling
wavefunction should not affect the final results, if the overall
symmetry is incorrect the method becomes very inefficient. 
In phase separated regions, we found that using the uniform
importance wavefunction resulted in larger statistical errors; 
unfortunately there is no easy way to construct a phase-separated 
trial function in this case.
This is another example of the great care that must 
be taken in choosing an appropriate trial function 
in projector methods \cite{laddercpmc}.
In Table I we compare energies calculated using the SSE and CPMC methods for
32 site systems with $U=1$ and $V=8$. For the rest the results in this paper,
we have used the SSE and IR methods exclusively, using SSE for
energies when possible and IR for structure factors and susceptibilities.

\setlength{\tabcolsep}{0.1in}
\begin{center}
\begin{tabular}{|l||l|l|l|} \hline \hline 
n & IR & SSE & CPMC \\ \hline
0.3125  & -0.461(2)  & -0.4644(2)  & -0.4656(2) \\
0.3750  & -0.451(2)  & -0.4484(2)  & -0.4506(2) \\
0.4375  & -0.382(3)  & -0.3825(1)  & -0.3829(2) \\
0.5000  & -0.293(3)  & -0.3000(2)  & -0.3002(3) \\
0.5625  & -0.211(3)  & -0.2105(2)  & -0.2108(4) \\
0.6250  & -0.113(3)  & -0.1179(2)  & -0.1168(6) \\
0.6875  & -0.025(4)  & -0.0261(2)  & -0.0200(6) \\
0.7500  & -0.070(3)  & 0.0629(3)  & 0.072(1) \\
0.8125  & 0.150(2)  & 0.1510(2)  & 0.1640(7) \\
0.8750  & 0.239(2)  & 0.2406(3)  & 0.245(1) \\ \hline
\end{tabular}
\end{center}
Table I: Comparison of QMC energies per site for $V=8$ and $U=1$. 
Statistical errors in last digit are
indicated in parentheses. The CPMC results used a free-electron
trial function and were scaled for $\Delta\tau\rightarrow 0$ 
from $\Delta\tau=0.1$ and $\Delta\tau=0.05$.

\section{Phase Separation}

\subsection{Phase Separation and Superconductivity}

We have already noted that an obvious motivation for mapping
carefully the regions phase separation in the extended Hubbard
model is that in a variety of 
strongly-correlated models, superconductivity has been found, or argued to
be present, in close proximity to phase separation boundary.
Such proximity is intuitively reasonable, 
since both superconductivity and phase separation result from
effectively attractive interactions. The case of the EHM 
for $V<0$ provides an illustration. In this region the model phase separates 
into a low-density phase and a high-density phase; the high-density
phase consists {\it either} of adjacent doubly occupied sites {\it or}
of adjacent single 
electrons, depending on the value of $U$.\cite{big2,big} Near 
these phase-separated regions at negative $V$ are well-established
regions of singlet and triplet superconducting fluctuations.

However, the possible proximity of superconductivity
to phase separation is also problematic,
for if one applies the Luttinger liquid
formalism and numerical techniques to calculate 
$K_{\rho}$ for relatively small systems in a region that is in fact
phase separated in
the thermodynamic limit, one will obtain incorrect results. Hence, before 
a mapping to Luttinger liquid parameters can be used, the
phase separation boundaries of 
a model should be accurately determined. To  quantify
this point, we here note that in one recent study of the EHM at quarter
filling \cite{pm} the phase separation boundary for $V\gg |U|$ was determined 
using a cluster Gutzwiller approximation to find the vanishing of the 
inverse compressibility. The {\it smallest} $V$ (for any of the
values of $U$ studied) for which phase separation was found 
was $V\sim 14$ (for $U \sim -2.5$). An exact diagonalization study gave 
comparable results.\cite{mz} Our QMC studies of larger systems 
reveal that phase separation occurs already at $V \approx 8$ at
quarter filling, in 
close proximity to where these previous studies indicated that dominant
superconducting fluctuations first appear.

\subsection{Strong--Coupling calculations in the $V\rightarrow\infty$ limit}

An effective model, first derived by Penc and Mila (PM), \cite{pm}
conveniently captures the {\it exact} behavior of the EHM for 
$V\rightarrow\infty$. For infinite $V$, any existing pairs
cannot be broken up (or moved), while single electrons cannot occupy 
neighboring sites and hence behave as spinless fermions. The effective 
model thus consists of immobile pairs and single spinless electrons.
For parameter values where pairs and unpaired electrons coexist, the 
minimum energy corresponds to having the pairs and the spinless fermions 
separated into two distinct regions, so as to minimize the kinetic 
energy of the spinless fermions by allowing them to move in the
largest possible region. The pairs are separated by one
lattice spacing. The energy for a given number of pairs and single 
electrons can then readily be calculated analytically in the thermodynamic
limit as the sum of $U$ times the number of pairs plus the energy of 
a spinless fermion chain:
\begin{equation}
E/N={m n U\over 2}-t {2\over \pi}(1-n) \sin[\pi{n\over 1-n}(1-m)]. 
\label{pmenergy}
\end{equation}
Here $m$ is the fraction of fermions forming pairs in the ground
state, $E/N$ is the energy per lattice site. The energy can be 
minimized with respect to $m$ to determine the ground state. For $U<-4$, 
one can show that all electrons are paired for all fillings, giving a spin 
gap and the boundary of the Luther-Emery region for $V\rightarrow\infty$. 
For $U > -4$, the ground state contains only unpaired electrons for
fillings less than a critical filling, above which the pair phase
starts to grow. At half filling, the system contains only pairs.
Equation (\ref{pmenergy}) thus provides an exactly solvable model 
exhibiting phase separation. 

While PM focused exclusively on the quarter-filled case, their energy 
expression (\ref{pmenergy}) can be used to obtain the phase-separated
region for 
all fillings $n$. If the system is phase separated with the 
high- and low-density phases at densities $n_{\rm HD}$ and $n_{\rm LD}$,
respectively, then phase separation occurs for the total (average)
densities $n$ in the range $n_{\rm LD} < n < n_{\rm HD}$.
In the thermodynamic limit the ground state energy must be linear as
a function of $n$ in this regime, reflecting the fact that adding
particles to the system only changes the relative sizes of the two
phases in a system with fixed particle number (canonical ensemble).

\begin{figure}
\centerline{\epsfig{width=3in,file=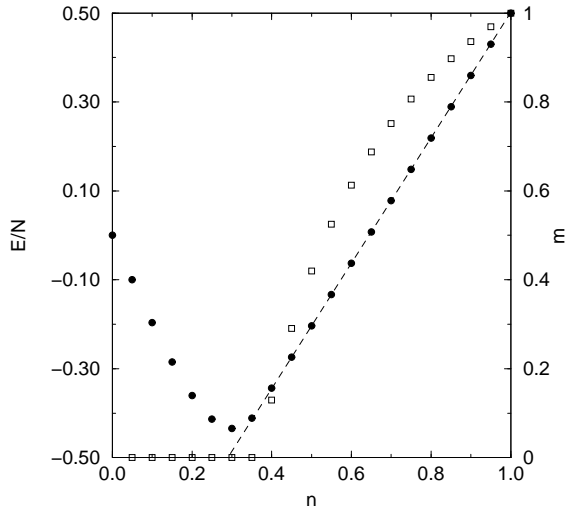}}
\caption{The exact energy per site for for $V\rightarrow\infty$ and
$U=1$ (solid circles, left axis) calculated using Eq.~(\ref{pmenergy}). 
The dependence on $n$ is completely linear for 
$n > 0.37$, as
indicated by the dashed line, reflecting phase separation. The open 
squares (right axis) denote the fraction $m$ of
fermions forming pairs in Eq.~(\ref{pmenergy}).}
\label{maxwell_infinite_v}
\end{figure}

\begin{figure}
\centerline{\epsfig{width=3in,file=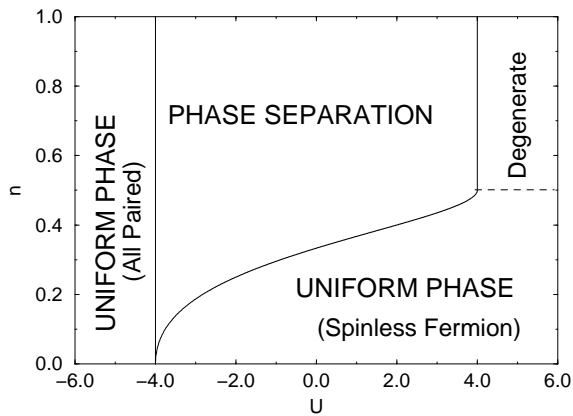}}
\caption{The region of phase separation (PS) (extending from $U= -4$
to $U = +4$) for infinite $V$. The other
phases are discussed in the text.}
\label{infinite_v_phase_diagram}
\end{figure}

Figure \ref{maxwell_infinite_v} shows the energy per site versus 
filling in the $V\rightarrow \infty$ limit. At low fillings the system
contains only un-paired electrons. Above a critical filling $n_{\rm LD}$, 
the energy becomes linear, reflecting phase separation as pairs are
formed. The linear behavior persists
all the way up to half filling ($n=1=n_{\rm HD}$), because the high density
phase consists of pairs occupying every other site. The phase separation 
density can be easily calculated as a function of $U$, resulting in the
phase diagram shown in Figure \ref{infinite_v_phase_diagram}. 
\noindent For $U>4$, the phase-separated and uniform states become degenerate at 
$V\rightarrow\infty$, which we label as non-phase separated because we
expect any finite $V$ will break 
the degeneracy and the resulting
state will not be phase separated. Mila and Zotos have shown that this
state in fact is uniform and insulating.\cite{mz} 
For finite $V$, we 
expect a smaller phase-separated region,
as pairs will be able to break up with a 
finite amount of energy. Thus we are able to confine our numerical 
investigation of phase separation to the range $-4 < U < 4$.

\subsection{Numerical Calculation of the PS Boundary}
Numerically, for finite $V$, the phase separation 
boundary can  be determined using the energy
as a function of the number of particles in the system, as discussed
above. However, a completely linear behavior will not be observed in a finite
system, because the boundaries between the high- and low-density phases 
then occupy a significant fraction of the lattice and raise
the energy per site by an amount that depends on the size of the
boundary regions (we use periodic systems and hence have two boundaries).
This will cause the energy as a function of $n$ to be concave, which
is not possible in the thermodynamic limit. A line can be drawn which
is tangent to the $E(n)$ curve at two points, which then constitute
estimates of the fillings $n_{\rm LD}$ and $n_{\rm HD}$ of 
the high- and low-density phases. In the absence of phase boundaries
(which become irrelevant to the energy per site in the thermodynamic
limit), the line corresponds to $E/N$ of a system separated into regions of 
densities $n_{\rm LD}$ and $n_{\rm HD}$. For small
system sizes, this ``Maxwell construction'' 
can be expected to be more accurate than signals of phase separation
based on, e.g., probability distributions of particle densities, since
the ground state energy typically exhibits only small finite-size effects in
cases where the thermodynamic limit state is uniform. It was previously
assumed that $n_{\rm HD}=1$ also for finite $V$.\cite{pm} As we will
see, in fact $n_{\rm HD} < 1$ in some parameter regions. In any case, we 
will refer to $n_{\rm LD}$ as the phase separation density.

As a complementary method of determining the phase separation boundary, as well
as to help in characterizing the phases, we have also used a criterion
based on the static charge structure factor $S_\rho(q)$:
\begin{equation}
S_\rho(q)={1\over N} \sum_{jl} e^{iq(j-l)} \langle n_j n_l - n^2\rangle .
\label{srho}
\end{equation}
Continuity at $q=0$ requires $S_\rho (q \to 0) = 0$ in a uniform system.
In a phase-separated periodic system of size $N$
in the canonical ensemble, $S_\rho (q)$
will have a maximum at the shortest non-zero wave-number $q=q_1=2\pi/N$, with
a divergence as $N \to \infty$. In small systems close to
the boundary of phase separation, this signal is, however, ambiguous,
since there will be a range
of parameters for which it is not possible to determine accurately the
$S (q \to 0)$ behavior based on the behavior for $q \ge q_1$. We
shall see examples of this in later sections.

Returning to the Maxwell construction, we note that
for it to be reliable one must be certain that 
the energy can be computed accurately.
Although there are no {\it a priori} approximations in two of
the QMC methods we use, near half filling at large $V$, we have noticed that 
the QMC simulations may converge poorly, apparently becoming stuck
in meta-stable states.  This occurs because the ground state
near half filling contains a significant
fraction of on--site pairs that require considerable energy to break 
up or move, making it difficult for the simulation to explore 
the full phase space of the model. The simulations are particularly hard 
when the ground state is phase separated. The system can then separate
into several alternating domains of high-and low-density phases, instead
of just one of each. The resulting large statistical errors
in the energy close to half 
filling cause problems in the determination of the tangent line and
thus the point of phase separation. Fortunately, exactly at $n=1$,
the system is always uniform
and the ground state energy can be calculated very accurately using exact
diagonalization, since the finite size effects are very small at this filling
(much smaller than for $n < 1$). This can be easily understood from 
perturbation theory around the static ($t=0$) ground state, which is 
{\it non-degenerate} at $n=1$ (and only there) and consists of alternating 
doubly occupied and empty sites (i.e., a period-two CDW). A simple 
second-order perturbation calculation gives an energy per site of 
\begin{equation}
E/N = {U\over 2} - {2 t^2\over3V-U} + \ldots .
\end{equation}
This thermodynamic expression compares remarkably well with exact 
diagonalizations on small systems; for example with $U=1$ and $V=10$, it 
gives $E/N=0.431034$, while the exact results for $N=6$ and $N=12$ are 
$E/N=0.431092$ and $E/N=0.431096$. Unfortunately, the smallness of the 
finite-size effects holds only exactly at half filling (since the particles
are no longer localized once the system is doped away from half filling). 

\begin{figure}
\centerline{\epsfig{width=3in,file=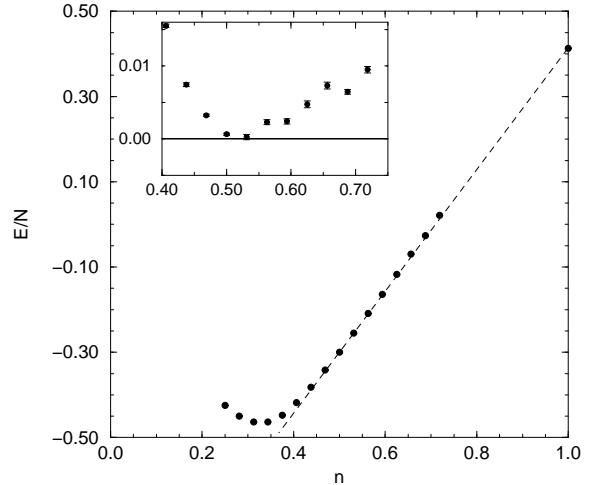}}
\caption{Maxwell construction for $V=8$ and $U=1$. The solid
circles are energies calculated for $N=64$. The dashed line goes
through the $n=1$ energy, and is tangent to the energy curve at 
the phase separation filling (density of the low-density phase)
$n_{\rm LD} \approx 0.53$. The inset shows the difference between 
the data points and the tangent line.}
\label{maxwell_u1}
\end{figure}

For $n \alt 0.7$, the QMC simulations converge well even for
phase-separated ground states
(for $V/t \le 10$), and we have been able to calculate $E/N$ to within 
absolute statistical errors of $\sim 10^{-4}-10^{-3}$ using the SSE
method (energies for small systems agree  with ED results). Figure
\ref{maxwell_u1} shows our Maxwell construction for $N=64$, $U=1$ and $V=8$.
Also shown is a plot of $\Delta E$, the difference between the $E/N$ values 
and the best fit tangent line. Using this method, we find that the system 
in this case phase separates at $n$ slightly above quarter filling. Data 
for $N=32$ give the same result, indicating that this is indeed close to 
the phase separation point in the thermodynamic limit. Thus,
for a quarter filled 
system, we can conclude that $V_{\rm PS}\approx 8.0$ for $U=1$
(since for $V=8$, phase separation occurs at $n$ only slightly above
$1/2$, and the critical filling decreases with increasing $V$). 
Previous studies at this filling\cite{mz,pm} found a $V_{\rm  PS} \approx 
14-18$ at $U=1$. Our phase separation
boundary is hence at significantly lower $V$ than 
reported and is very close to the $K_\rho=1$ curve obtained
using exact diagonalization.\cite{mz,pm} This is a clear indication of 
possible problems with the prediction of a novel superconducting region
for $V \gg |U|$.

\begin{figure}
\centerline{\epsfig{width=3.5in,file=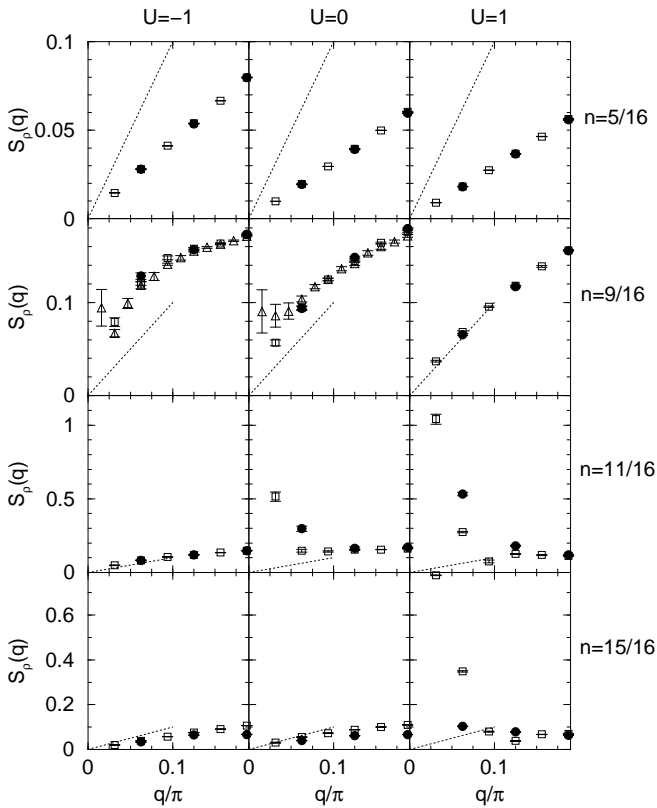}}
\caption{The static charge structure factor $S_\rho(q)$ for small
momenta $q/\pi$ at various fillings for $V=8$ and $U=-1,0,1$.
Filled circles are for $N=32$, open squares for $N=64$ and
triangles for $N=128$. As explained in Section IV,
the straight lines all have slope = $\pi$, 
corresponding to the $q \to 0$ behavior if the LL exponent 
$K_\rho =1$.}
\label{srho_u1u0}
\end{figure}

A definite disadvantage of using the Maxwell construction method for 
determining phase separation
boundaries is that the need to calculate very precise 
energies causes this approach to be extremely time consuming. Accordingly,
as already noted, we employed also a second method of detecting
phase separation, 
which relies on the behavior of the static charge structure factor
defined in Eqn. (\ref{srho}) for $q\rightarrow 0$. 
Figure \ref{srho_u1u0} shows $S_\rho(q)$ for several densities at $U=-1,0,1$,
and $V=8$. For $U=1$ we obtain phase separation
at a density $n \approx 0.59$  from
$S_\rho(q)$, compared to $n \approx 0.53$ from the Maxwell construction. 
Note, however, that at $U=1$ and $n=9/16$, $S_\rho(q)$ shows no
phase separation, whereas the Maxwell
construction implies the system is already phase separated. $S_\rho(q)$ 
gives a slightly smaller phase-separated
region than the Maxwell construction because a 
clear signal of an increase as $q \to q_1$ for a relatively small lattice 
requires that the system already be well inside the region
of phase separation.

Examining the region of phase separation for $U\le 0$
reveals several interesting properties:
whereas at $U=1$ the system stays phase separated for all densities $n < 1$ 
above some critical density $n_{\rm LD}$, for $U=0$ we see clear signs that
phase separation {\it disappears} at high densities. This can be seen in the 
structure factors in Figure \ref{srho_u1u0}, which go smoothly to zero for 
$U=-1,0$ and $n>0.9$. The Maxwell construction is also consistent with 
phase separation
to a state with a high-density phase at filling less than $n=1$, although
the relatively large statistical errors in the energies close to half filling
make it difficult to obtain an independent estimate of $n_{\rm HD}$.
However, investigating the charge structure factor for several system sizes,
it is clear that the uniform phase for $n \alt 1$ is stable. The peak position
of the charge susceptibility is also consistent with a high-density phase 
different from the $U=1$ case (see next section). We conclude that there is 
a ``re-entrant'' uniform state for $U\sim 0$. In addition, from simulations 
carried out at larger $V$ ($V=10$), we see the phase-separated
region moving further into
the $U<0$ side of the phase diagram and the re-entrant uniform state
remains. As we will discuss in the next Section, the re-entrant state
has strong $2k_{\rm F}$ CDW fluctuations. 

We believe that the the re-entrance of the uniform state for $U \alt
0$, but not for $U > 0$, can be qualitatively understood in the
following way: When $V$ is large and $n \to 1$, pairs are formed to
avoid the nearest-neighbor repulsions. For $U > 0$, the pairs have
positive energy. 
\noindent Via phase separation, the total energy is minimized
by balancing the potential energy of the pairs (which increases with
the number of pairs) and the kinetic energy of a low-density phase
with dominantly unpaired electrons. The pressure of the unpaired
phase forces the distance between the pairs in the immobile high-density phase
to be minimized 
to one lattice spacing (the kinetic energy of a pair is 
very small due to the $V$ barrier to hopping, and the motion
of the pairs can be be neglected).
The combined effects of avoiding the repulsive on-site $U$ 
and delocalization of mobile unpaired
electrons always favor phase separation as $n \to 1$ if $V$ is large.

On the other hand, for $U < 0$,
the pairs have negative energy and maximizing their
number is favorable. For $n$ close to $1$ a uniform state with mobile 
pairs can then be stabilized if $V$ is not too large. However, as
the density is lowered the number of negative-energy  pairs decreases,
and if $|U|$ is small a phase separated
state can then again be lower in energy due 
to the lower kinetic energy of dominantly unpaired electrons in the
low-density phase. Increasing $|U|$ ($U < 0$), maximizing the number
of mobile pairs becomes more favorable, and the size of the region
of phase separation decreases, in agreement with our results.

For $U>1$ at $V=8$, we find that the phase separation boundary moves towards 
higher fillings, becoming hard to discern for $U\sim 4$, as expected from 
the infinite $V$ solution. For $U \alt -1$, based on 32-, 64- and limited 
128-site data, phase separation does occur for $V=8$, but is much harder 
to detect than at $U=1$ and $U=0$. Increasing $V$ to $10$, we cannot obtain 
as accurate results as for $V=8$, but we do find clear signs that the phase
separation region moves to lower values of $n$, as expected if there
is a continuous
evolution of the phase separation
boundary to the infinite $V$ solution discussed in the
previous section. We also find that phase separation does not occur at all 
for $V=4$, and for $V=6$ appears to be present only in a small region for 
$U > 0$. 

Finally, we stress the need to use more than 
one system size in applying the approach based on $S_\rho(q)$,
as otherwise the analysis
may give ambiguous or erroneous results for the phase separation boundary.
For instance, based
on the 32- and 64-site results shown in Fig. \ref{srho_u1u0},
there would appear to be no
phase separation at $U=-1$ and
the slope of $S_\rho (q)$ versus $q$ as $q \to 0$
exceeds $\pi$ for some fillings. 
This would imply (see next section) that $K_\rho>1$ and therefore dominant 
superconducting
correlations in this region. However, based on data for larger ($N=128$)
lattices, it is likely that the system is phase separated at
these fillings (see $n=9/16$ for both $U=-1$ and $U=0$ in Figure 
\ref{srho_u1u0}).

\section{Correlations and Fluctuations}

\subsection{Methods to calculate LL parameters}

In this section we describe results of calculations for a variety of
correlation functions of the ground state of the EHM in
{\it non}-phase-separated regions 
using various different estimators for the Luttinger liquid
correlation exponents 
$K_\rho$ and $K_\sigma$. For models (like the EHM) with spin--rotation 
symmetry, the exponent $K_\sigma=1$ \cite{voit} {\it except} in the 
Luther--Emery phase, in which case only $K_\rho$ has meaning, since the 
system is spin-gapped and not a LL in the spin sector. For $K_\sigma=1$, 
the asymptotic equal-time correlation functions in a LL have simple
power law forms governed by $K_\rho$:

\begin{eqnarray}
N_{\rm SDW,CDW}(r) &\sim & r^{-(1+K_\rho)}\cos(2k_F r), \\
N_{\rm SS,TS}(r) &\sim & r^{-(1+1/K_\rho)}. 
\end{eqnarray}

Hence, when $K_\rho<1$, SDW/CDW correlations dominate at large $r$, while
for $K_\rho>1$, superconducting correlations dominate. At weak coupling,
the renormalization group ``g--ology'' procedure can be used to
determine the exponents $K_\rho$ and $K_\sigma$.\cite{solyom,emery}
Away from weak coupling, one must use the Luttinger liquid
equations that relate
$K_\rho$ to other physical observables of the model. In our present 
notation, the relations we use are \cite{schulz}
\begin{mathletters}
\begin{eqnarray}
K_\rho & = & D_\rho/ 2v_{\rho}, \label{krhoa} \\
K_\rho & = & \pi v_{\rho} \kappa / 2, \label{krhob}  \\
K_\rho & = & (\pi \kappa D_\rho / 4)^{1/2} \label{krhoc},
\end{eqnarray}
\label{krho}
\end{mathletters}

\noindent
where $D_\rho$ is the Drude weight of the optical conductivity (the charge
stiffness), $v_{\rho}$ the velocity of the charge excitations,
and $\kappa$ the compressibility.  The use of three different relations 
for $K_\rho$ is important to verify the validity the calculation. 
In particular, the three relations will give inconsistent results
if the system is not a LL, or if finite--size or other systematic
errors are present. 

The Drude weight can be calculated from QMC simulations as 
the $\beta \rightarrow \infty$ limit of \cite{kohn,szw} \\ 
\begin{equation}
D_\rho =\pi[\langle -K/N \rangle - \Lambda_c(i 2\pi/\beta)],
\end{equation}
where $K$ is the kinetic energy, and $\Lambda_c(i\omega_m)$ 
is the Matsubara frequency--dependent charge current--current correlation 
function:
\begin{equation}
\Lambda_\rho(i\omega_m)={1\over N}\int_0^\beta d\tau
e^{i\omega_m \tau} \langle j(\tau) j(0)\rangle .
\end{equation}
The standard method used in exact diagonalization calculations of the
compressibility $\kappa$ employs a finite difference approximation involving
ground state energies for different numbers of particles:\cite{pssc1}
explicitly,
\begin{equation}
\kappa^{-1} = N_e {[E(N,N_e+2)+E(N,N_e-2)-2E(N,N_e)]\over 2},
\label{kappa_eqn}
\end{equation}
where $E(N,N_e)$ is the ground state energy of $N_e$ electrons on an
$N$-site lattice. The 
compressibility may be also computed from the $q \rightarrow 0$ limit of 
the static charge susceptibility; 
\begin{equation}
\kappa = \chi_c(q\to 0),
\label{kappax}
\end{equation}
 where
\begin{equation}
\chi_\rho(q)  ={1\over N} \sum_{j,l}e^{iq(j-l)}\int_0^\beta
d\tau \langle n_j(\tau) n_l(0)\rangle ,
\label{chiq}
\end{equation}
where $n_j(\tau)$ is the charge at site $j$ and imaginary time $\tau$.
This definition avoids the errors due to discretization in the 
finite-difference definition (\ref{kappa_eqn}), and we will therefore
primarily use Eq.~(\ref{kappax}) here. Finally, the 
velocity $v_\alpha$ associated with the gapless charge ($\alpha=\rho$) or 
spin ($\alpha=\sigma$) mode may be extracted from the ratio \cite{cuo2}
\begin{equation}
W_\alpha(q) = 2 S_\alpha(q)/\chi_\alpha(q),
\label{bound}
\end{equation}
where $S_\alpha(q)$ is the static structure factor given by
Eq. (3). $W_\alpha(q)$ gives an upper bound to the energy of the lowest 
excitation of momentum $q$ and becomes the exact excitation 
energy as $q \to 0$ in a LL, so that $v_\alpha$ can be directly extracted 
from the $q$-dependence for small $q$. Hence, the quantities needed for 
all three estimates of $K_\rho$ defined in Eqs.~(\ref{krho}) can be 
computed directly from QMC data. Examining Eqs.~(\ref{krhob}), (\ref{kappax}),
and (\ref{bound}), one can see that $K_\alpha$ is also given directly by
the slope of the static structure factor as $q\to 0$:
\begin{equation}
K_{\rho,\sigma}={1\over \pi q} S_{\rho,\sigma}(q\rightarrow 0).
\label{krho_slope}
\end{equation}
This relation may also be obtained directly from the calculation
of the charge-charge or spin-spin correlation functions in LL
theory:\cite{schulz}
\begin{equation}
\langle n_{\alpha 0} n_{\alpha r} \rangle = -{K_\alpha \over (\pi r)^2} +
{\cos(2k_F r)\over r^{1+K_\alpha}} + \cdots
\label{ll_correlation}
\end{equation}
The Fourier transform of the non-oscillating term
of equation (\ref{ll_correlation}) leads to the expression 
(\ref{krho_slope})
for $K_\rho$ and $K_\sigma$. As the structure factor is usually
much easier to calculate numerically than the compressibility or
the Drude weight, Eq.~(\ref{krho_slope}) is quite useful for
calculating Luttinger liquid
exponents. In particular, because $S_\alpha(q)$ only depends on
equal-time correlations in imaginary time, for most QMC
methods the structure factors converge much faster than
$D_\rho$ or the susceptibilities and hence may provide a better
estimate for $K_\rho$. But the important caveat applies that 
the consistency among all three relations must be checked, which
requires more detailed calculations. Moreover, one additional 
consistency check is available in regions where one expects no
spin gap. There it is required that $K_\sigma=1$, which can be
verified (or disproved!) from the $q \rightarrow 0$ limit of $S_\sigma(q)$.
Of course, the limit $q\rightarrow 0$ is impossible to attain
strictly in numerical simulations of finite systems.
Since Eq.~(\ref{bound}) is an upper bound of the lowest 
excitation energy at momentum $q$, we expect that values of $K_\rho$
and $K_\sigma$ calculated from equation (\ref{krho_slope}) to be in general 
slightly {\it larger} than their true values. Effects of non-linearity in the true 
lowest excitation energy should be smaller for typical
values of the smallest $q$ accessible (i.e., we expect effect of the 
broadening of the mode to be larger than effects of non-linearity).

One additional complication to using the slope of the structure factor
to determine $K_\rho$ is that for very small fillings, $2k_F$ may be 
close in momentum to the smallest $q=q_1$. A strong broad CDW peak at
this momentum may then affect the behavior all the way down to $q_1$. 
The value obtained for $K_\rho$ will then still be an upper bound,
but may be much larger than the true value.

Since from Eq. (\ref{krhob}) one sees that $K_\rho $ is proportional to the
compressibility, $\kappa$, which (naively) should diverge at the phase
separation boundary, calculations of $K_\rho$ using Eq.~(\ref{krhob}) 
or equivalently Eq.~(\ref{krho_slope}) near a region
of phase separation typically show a strongly increasing $K_\rho$. This
increase of $K_\rho$ has often been interpreted (in a variety of 
one--dimensional models) as evidence for a region of dominant
superconducting correlations
\cite{pm,so,pssc1,pssc2,dagotto}. A more careful analysis begins by noting 
that in an infinite system, $\kappa$ jumps {\it discontinuously} from a finite
value to infinity at the phase separation
boundary if the transition is first order, which 
is normally the case. However, as observed above in Sec.~II, and
also recently by Hellberg and Manousakis,\cite{steve} $\kappa$ in a finite 
system can diverge only {\it inside} the region where the infinite
system is phase
separated, due to the energy cost of the interface between the two phases. 
Hence, a result $K_\rho > 1$ based on a diverging $\kappa$ obtained from
small systems may be misleading, since it is possible that the system is
already phase separated in the thermodynamic limit. 

As already noted,
all these considerations raise legitimate concerns about the reliability 
of the previous calculations of the phase separation
boundary in the EHM and hence also 
lead to concerns about whether $K_\rho$ exceeds unity in the uniform phase.
Indeed, a recent QMC study  \cite{cuo1,cuo2} of a two-band 1D Hubbard model 
related to the EHM (a two-band ``Cu-O'' chain) showed that a first-order 
phase separation
transition preempts superconductivity in all but a small part of the 
phase diagram, in contrast to a previous ED study, \cite{pssc2} which 
indicated that superconducting
fluctuations always dominate in the neighborhood of phase
separation. 

\subsection{QMC results for $K_\rho$ in the EHM}

Using QMC techniques to evaluate $K_\rho$ accurately via all three 
Eqs.~(\ref{krho}) is extremely time consuming, as $D_\rho$ in particular is 
difficult to calculate in the $V\gg U$ parameter region. Accordingly, in this 
section we first present results using all three relations (\ref{krho}) 
for only one value of $U$, and then use these detailed results to 
``benchmark'' our alternate method of getting $K_\rho$ directly from 
Eq.~(\ref{krho_slope}). We choose $U=1$ for these calculations, and find 
excellent agreement between the two methods, which then justifies our
use of  the static structure factor method for other values 
of $U$.

In addition to the QMC simulations for 32-, 64-, and 128-site chains, we have 
performed Lancz\"os ED calculations for 16 sites, in order to investigate 
systematically the finite-size effects. We have also checked the QMC
simulations against the ED data for this system size. In our ED calculations,
we use Eq.~(\ref{kappa_eqn}) to define the compressibility $\kappa$
and to extract
the charge velocity from the lowest charge excitation energy.

\begin{figure}
\centerline{\epsfig{width=3in,file=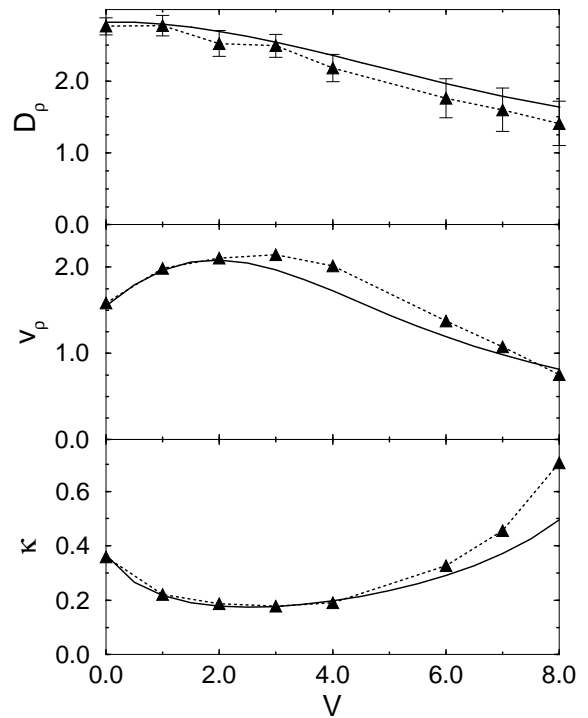}}
\caption{Comparison of the Drude weight ($D_\rho$), the charge
velocity ($v_{\rho})$, and the compressibility ($\kappa$) for $U=1$ as 
computed by ED with $N=16$ (solid curves) and QMC with $N=64$ (triangles).
Where not shown, the QMC error bars are approximately the size of the 
symbols (somewhat larger for $V=7,8$). The dashed lines provided
guides to the eye for the QMC data.}
\label{dvk}
\end{figure}

We begin with a comparison (for $U=1$ and varying $V$) of the various 
low-energy parameters used in calculating $K_\rho$. Figure \ref{dvk} 
compares the Drude weight, the charge velocity, and the compressibility 
as computed for $n=1/2$ (quarter filling). 
Up to $V \approx 2$ the ED and QMC
results agree almost perfectly. For larger $V$ the deviations are due
to the larger finite-size effects in the ED data. For $V \alt 6$,
the finite-size errors in the ED results seem to be greatest for 
$v_\rho$; hence we expect that Eq.~(\ref{krhoc}), which is the only
Luttinger liquid relation
{\it not} involving $v_\rho$, will give the best estimate for $K_\rho$ 
in small systems. For large $V$, the slower increase of the compressibility
as computed by ED is likely primarily due to discretization errors
in Eq.~\ref{kappa_eqn}, which become large in regions where the energy
curvature changes rapidly as a function of $N_e$ (as is the case close to
phase separation).

\begin{figure}
\centerline{\epsfig{width=3in,file=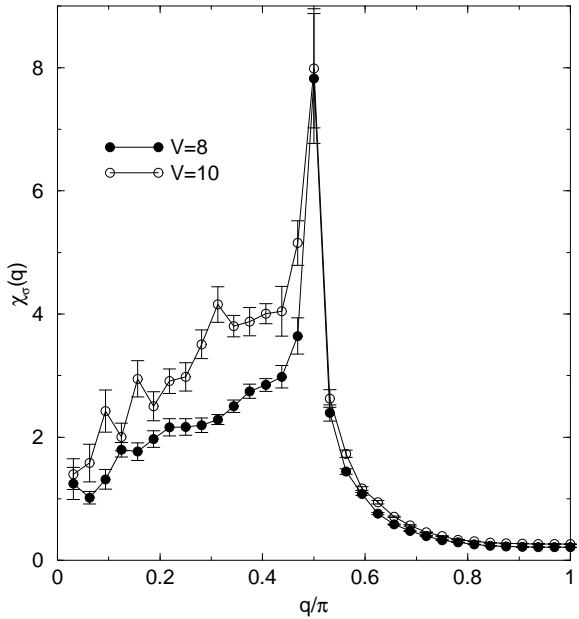}}
\caption{$\chi_\sigma$ for $U=1$ and $n=0.5$ (quarter-filling) at
$V=8$ (solid circles) and $V=10$ (open circles).}
\label{figchis}
\end{figure}

\begin{figure}
\centerline{\epsfig{width=3in,file=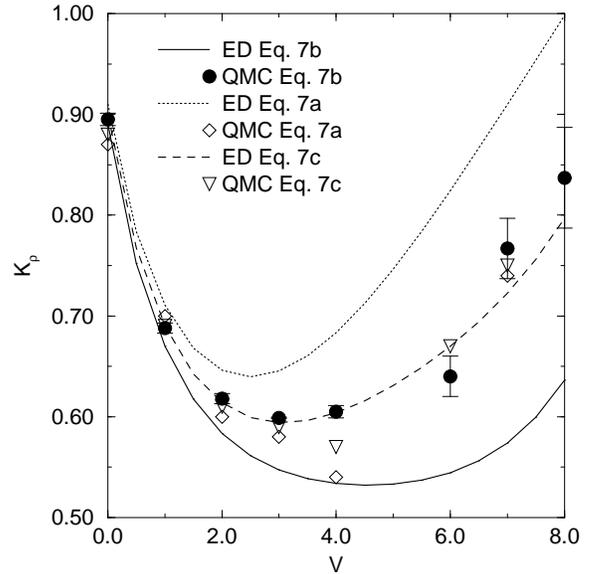}}
\caption{$K_\rho$ vs. $V$ for $U=1$ calculated using all 
Eqs.~(\protect{\ref{krho}}). ED results (curves) for 16 sites are
compared with QMC results for 64 sites (symbols). Error bars are 
shown only for Eq.~(7b) for clarity; error bars for Eqs.~(\protect{\ref{krhoa}})
and (\protect{\ref{krhoc}}) are larger. Phase separation for this filling 
and $U$ occurs at $V\approx 8$}
\label{k_rho_fig}
\end{figure}

Figure \ref{k_rho_fig} shows results for $K_\rho$ as a function of $V$ 
calculated for $U=1$ by Lancz\"os diagonalization and QMC simulations. 
The data points based on Eq.~(\ref{krhob}) are equivalent to 
using {\it only} the structure factor in determining $K_\rho$
according to Eq.~(\ref{krho_slope}). The error
bars of the other methods are not shown for clarity; in general they are 
at least twice as large as the errors using the structure factor slope 
method. Further, in Figure \ref{k_rho_fig}, the $K_\rho$ values for $V=8$ 
calculated using Eqs.~(\ref{krhoa}) and (\ref{krhoc}) have been left off, 
as their statistical errors are too large for this large $V$. Given the 
equivalence of using Eq.~(\ref{krhob}) to the calculation of the slope of the
structure factor at $q =0$, it is clear that the slope method provides a 
good estimate for $K_\rho$, as it agrees with the other methods within 
error bars but shows smaller statistical fluctuations. Previous $K_\rho$ 
calculations \cite{pm,big} using ED methods had seen large differences 
among the different relations (\ref{krho}).
As seen in Fig.~\ref{k_rho_fig}, these differences are greatly reduced
using larger lattices. Indeed, our QMC results on 64-site systems show 
that the three relations give equivalent results, to within error bars. 
Most importantly, all the $K_\rho$ values obtained for $V<8$ and $U=1$ 
are less than one. The Maxwell construction, shown in
Fig.~\ref{maxwell_u1}, indicates that phase separation occurs at quarter
filling for $V$ only slightly larger than $V=8$. 
It is therefore clear that there is no extended region where $K_\rho > 1$ 
before phase separation, although there is a definite increase as the phase 
separation boundary is approached. We can of course not strictly rule out 
the existence of an extremely narrow region where $K_\rho$ exceeds one.
However, this seems unlikely in view of our results for the spin
susceptibility at $V=8$ and $10$ shown in Figure \ref{figchis}. In
both cases there is a clear peak at $q=2k_{\rm F}$, which would not
be expected if $K_\rho$ exceeds unity. Since our Maxwell construction
indicates that the $V=10$
\noindent system should be phase separated in the 
thermodynamic limit, we can conclude that the dominant SDW fluctuations
persist at the phase separation boundary and there is no region with
$K_\rho > 1$.

For other fillings and $U$ we have concentrated on $V=8$. For $1<U<4$ we find
similar behavior in $K_\rho$ (based on the slope of $S_\rho(q)$ for 32 
and 64 site systems) to the $U=1$ case: $K_\rho$ increases as the density 
increases, but does not exceed one before phase separation. For $-3<U< 0$, our
determination of the phase separation
boundary is not as accurate, and in some cases
calculations for for 32- and 64-site systems give $K_\rho\agt 1$ in regions
where there are no clear signals of phase separation. However, in
cases where we have also carried out calculations for $N=128$, the
general trend seems to favor phase separation over a stable uniform
state with $K_\rho > 1$ (recall the data for $U=-1$ and $U=0$ at $n=9/16$
in Fig.~\ref{srho_u1u0}). 

The persistence of finite-size effects in
calculations of $V_{\rm PS}$ and $K_\rho$ close to phase separation
for systems as large as $N=64$ sites emphasizes the importance of 
studying and understanding finite-size effects in calculations of 
$K_\rho$ and related quantities. In Sec.~V we will discuss the
complete phase diagram of the model and comment further on the
behavior of $K_\rho$ as $V \to V_{\rm PS}$.

\subsection{Unusual charge correlations in the EHM}

Luttinger liquid
theory predicts structure in the charge or spin response {\it only}
for multiples of $2 k_F$, which is a consequence of the low-energy
effective model being linearized about $\pm k_F$. Peaks are typically 
seen only at $q=2k_{\rm F}$ and/or $q=4k_{\rm F}$ (only for the charge
if there is a spin gap). The $2k_{\rm F}$ peaks should diverge as
$T\to 0$ if $K_\rho < 1$, and $4k_{\rm F}$ peaks are divergent
for small $K_\rho$.\cite{schulz} Divergences in finite systems can of
course be seen only as $N \to \infty$.

In studying the $V\gg U$ region of the EHM, we have found strong 
charge response peaks also at wavevectors that are not multiples
of $2 k_F$; we shall use the term ``anomalous peaks'' to refer to these unusual
charge correlations. Importantly, however, these anomalous peaks appear to be
associated with gapped modes. As the temperature decreases and $N$
increases they do {\it not} diverge. Hence we believe the system is
still a LL in this region for sufficiently low energies. We
shall comment further below on the interpretation of these anomalous peaks.

\begin{figure}
\centerline{\epsfig{width=3.8in,file=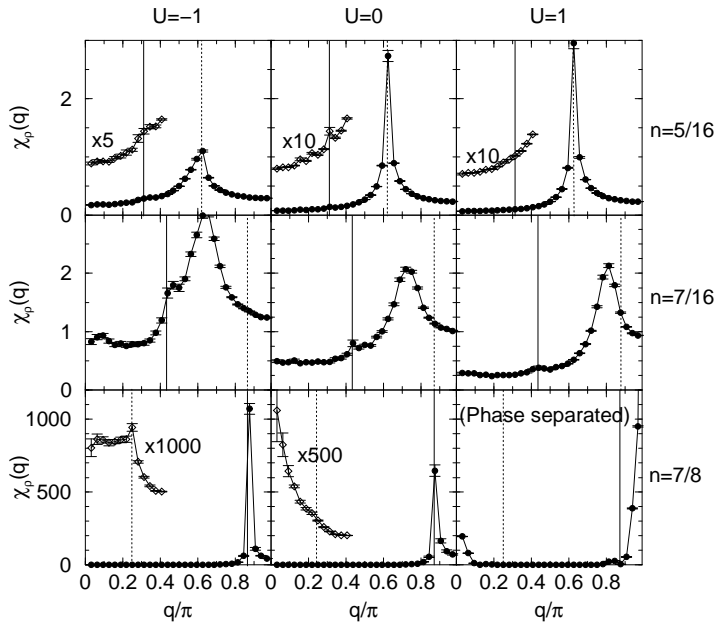}}
\caption{Charge susceptibility for $V=8$, $U=-1,0,1$. All data are for 64
site systems. The vertical solid and dashed lines indicate $q=2k_{\rm  F}$ 
and $q=4k_{\rm F}$, respectively.  Note that for $n>1/2$, $4k_F=\pi (2-2n)$ 
in the reduced zone scheme. Lines between data points are guides to
the eye.} 
\label{v8u0_corr}
\end{figure}

Figure \ref{v8u0_corr} shows the charge susceptibility for several
fillings for $V=8$ and $U=-1,0,1$. Starting with the $U=1$ data, we see 
that at small fillings the susceptibility is dominated by a large peak at 
$q=4k_F$. This corresponds to a system with almost no pairs, where
the particles behave essentially as spinless fermions. For higher
fillings the large peak shifts from the value of $4k_{\rm F}$ to a 
slightly {\it lower} momentum --- which for later purposes we call 
$4k^{\rm eff}_{\rm F}$ --- whose value depends on the filling as well
as the other model parameters. As one moves further into the phase-separated
\noindent region (above $n \approx 5/8$) a strong peak develops at $q=\pi$, 
independent of $n$ (and hence $k_F$). This corresponds to the 
wave-vector of the high-density phase in the phase-separated region, which is 
a CDW state in which double occupancies alternate with empty sites.

Turning to the cases for $U=-1$ and $U=0$, the behavior is quite similar
for low and intermediate densities. As in the $U=1$ case, at low fillings
one sees a large $4k_{\rm F}$ peak and a much weaker (but still
observable in the $U=0$ data) $2k_F$ peak. For intermediate fillings
near the phase separation
boundary, the main peak is no longer at $4k_{\rm F}$ but rather
at $4k^{\rm eff}_{\rm F}$ as for $U=1$. This is shown clearly in the
$U=0$ data for $n=7/16$, which is outside of the phase-separated region
($n < n_{\rm LD}$). At large fillings ($n > n_{\rm HD}$), instead of the 
peak at $q=\pi$ seen in the $U=1$ data, the $U=-1$ data shows a strong 
peak at $2k_{\rm F}$, reflecting the re-entrance (as a function of $n$) 
of the homogeneous phase, which is characterized by $2k_{\rm F}$
CDW fluctuations. A weaker $4k_{\rm F}$ peak can also be seen at this filling.
For $U=0$, $n=7/8$ is very close to our estimated re-entrance point 
($n_{\rm HD} \approx 0.88$). 

\begin{figure}
\centerline{\epsfig{bbllx=100pt,bblly=185pt,bburx=450pt,bbury=590pt,
        width=3in,file=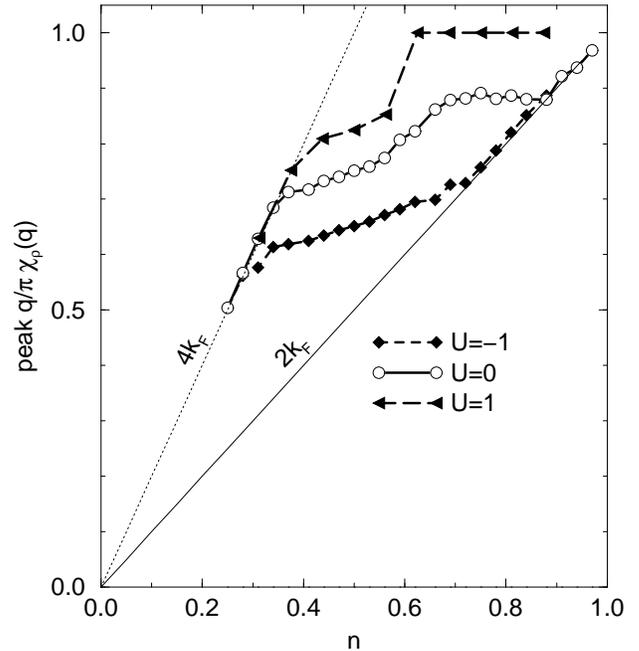}}
\caption{Momentum value of the largest peak in the
charge susceptibility $\chi_\rho(q)$ for a 64-site system
as a function of band filling $n$ for $V=8$, $U=1,0,-1$.
The peak position was determined by fitting a 2$^{nd}$
degree polynomial to points near the peak. The solid and dotted lines indicate
$q=2 k_F$ and $q=4 k_F$ respectively.}
\label{v8u0_peak}
\end{figure}

Figure \ref{v8u0_peak} shows the location of the dominant peak in the
charge susceptibility as a function of filling for $V=8$ and $U=-1,0,1$.
For small fillings, all the systems are dominated by the $4k_F$ charge 
response. As the filling is increased, the anomalous 
$4k^{\rm eff}_{\rm F}$ peak dominates the response for these 
system sizes. For still higher fillings the normal $2k_{\rm F}$ peak
dominates. As already noted, we see a $2k_{\rm F}$ peak also at the
densities where the anomalous peak is present 
(see Figs.~\ref{v8u0_corr} and \ref{v8u0_peak}), and we expect it to
\noindent diverge as $N \to \infty$ and $T \to 0$, reflecting asymptotic
LL behavior. Note the constant behavior of the peak position for $U=0$ 
at $0.65 \alt n \alt 0.88$; this indicates that this peak originates from 
the high-density phase of the phase-separated system, which apart from 
growing in size as $n$ increases remains unchanged in the phase
separation region. 
This behavior is seen less clearly for $U=-1$,
where the region of phase separation
is apparently very narrow. Close to half filling
a uniform state again stabilizes for $U=-1,0$, whereas the $U=1$ system 
remains phase separated all the way up to half filling.

\begin{figure}
\centerline{\epsfig{width=3in,file=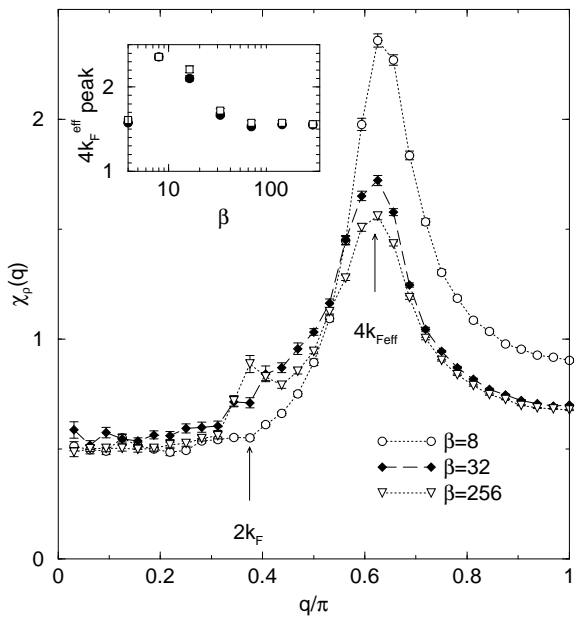}}
\caption{Behavior of non--LL ``anomalous'' $4k^{\rm eff}_{\rm F}$ 
charge peak in $\chi_\rho(q)$ for $V=8$, $U=-1$, $n=3/8$ vs. 
temperature. The symbols connected by dashed lines in the main figure
are for 64-site systems. The inset shows the peak value vs. $\beta$ for 
32-site (circles) and 64-site (squares) systems. Note also the presence 
of a normal $2k_{\rm F}$ peaks which grows with increasing $\beta$}
\label{non_divergent}
\end{figure}

Since the anomalous charge response is not at a wavevector
possible within the Luttinger liquid formalism, it is important to determine
whether the corresponding $4 k^{\rm eff}_{\rm F}$ peak diverges with
decreasing temperature and thus has thermodynamic relevance in the
low-energy sector. Figure \ref{non_divergent} shows the momentum
dependence of the charge susceptibility at three different temperatures
for a parameter set where the anomalous charge fluctuations are strong.
The inset shows the peak value versus inverse temperature for two
different system sizes. We see that as the temperature is lowered, the 
anomalous peak does not diverge (it in fact has a maximum at a finite
temperature), and there is almost no dependence on $N$. 
\noindent Nonetheless, this
non-LL charge peak is still relatively low in energy. Using equation 
(\ref{bound}) we are able to estimate an upper bound of $E\sim 0.75 t$ 
above the ground state at $V=8$, $U=-1$ and $n=3/8$. In
Figure \ref{non_divergent}, one clearly sees
that the normal $2k_F$ peak indeed grows with
decreasing temperature, as expected in a Luttinger liquid
with $K_\rho < 1$,
 although the amplitude is relatively weak even at $\beta=256$.

We believe that the anomalous charge response is due to on-site pairs
that are sufficiently long-lived (due to the difficulty of breaking them
up via high-energy intermediate states) to form a CDW together with the
single particles of the system. Specifically, in the infinite $V$ effective 
model of Penc and Mila,\cite{pm} there are two kinds of particles; spinless 
fermions and bosons with charge two.
The number of doubly occupied sites in the simulation of the EHM corresponds
to the number of bosons, and hence we can calculate the average
{\it total} number
of particles (fermions {\it plus} bosons) within the 
effective model. We find that the momentum of the anomalous charge peak is 
consistent with spacing uniformly a number of particles equal to this
total the number of particles. Hence, there is a tendency to form 
a CDW consisting of a mixture of fermions and bosons which repel each
other. Nevertheless, the
non-divergence of the anomalous peak, the presence of a $2k_F$ peak,
and the consistency among the Luttinger liquid
relations indicate that the asymptotic 
low-energy properties are still those of spin-1/2 fermions forming a LL.

\subsection{Spin gap in the EHM}

Our results for the spin susceptibility of the extended
Hubbard model in the $V\gg |U|$ region show either  a
\begin{figure}
\centerline{\epsfig{width=3.2in,file=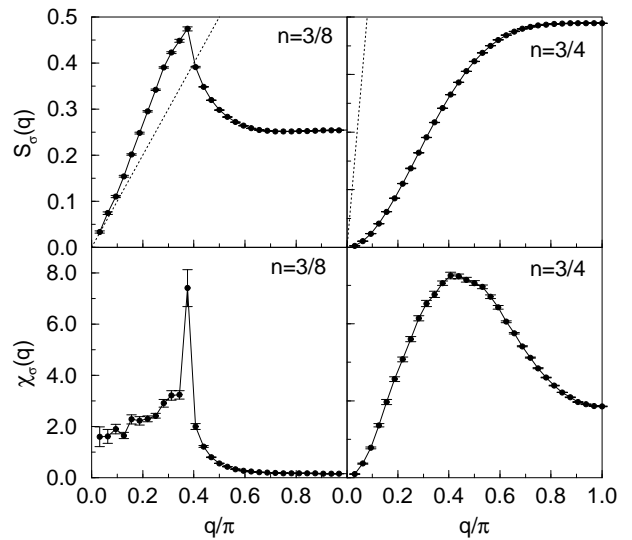}}
\caption{$\chi_\sigma(q)$ and $S_\sigma(q)$ comparing spin-gapped 
to non-gapped cases. 
Left panels are for $V=8$, $U=-1$, and $n=0.375$, where no gap is
present. Right panels are for same $U$ and $V$ but $n=0.75$, where the system
is gapped. Solid curves connecting data points are guides to the eye. 
The dotted lines indicate slope $\pi$ for $S_\sigma(q)$ vs. $q$.}
\label{spin_gap}
\end{figure}%
\noindent normal Luttinger liquid $2 k_F$ peak
or the presence of a spin gap. In
general, we find that the spin response is much noisier and harder to
converge numerically than the charge response, as expected in a parameter
region dominated by charge correlations. 
To determine whether the 
system is spin gapped, we examine the spin susceptibility $\chi_\sigma(q)$
in the limit $q\rightarrow 0$. If $\chi_\sigma(q\rightarrow 0)=0$, the system
is spin gapped. The presence of a spin gap can also be inferred from
the structure factor $S_\sigma(q)$: If there is no spin gap,
then $K_\sigma=1$, which translated into a slope $\pi$ for the
structure factor versus $q$
(as discussed in Sec.~IVA). In a spin-gapped system 
$S_\sigma(q)$ should approach zero faster than linearly and in particular
should fall below the line with slope $\pi$ (instead of approaching
this line from above, as expected if there is no spin gap). Again, one must 
use care in phase-separated
regions, as the $\chi_\sigma(q)$ will show a mixture of 
responses from both phases. Figure \ref{spin_gap} shows examples of 
$\chi_\sigma(q)$ and $S_\sigma(q)$ for parameters with and without a 
spin gap. The two different behaviors discussed above can be clearly
discerned.

\begin{figure}
\centerline{\epsfig{width=3in,file=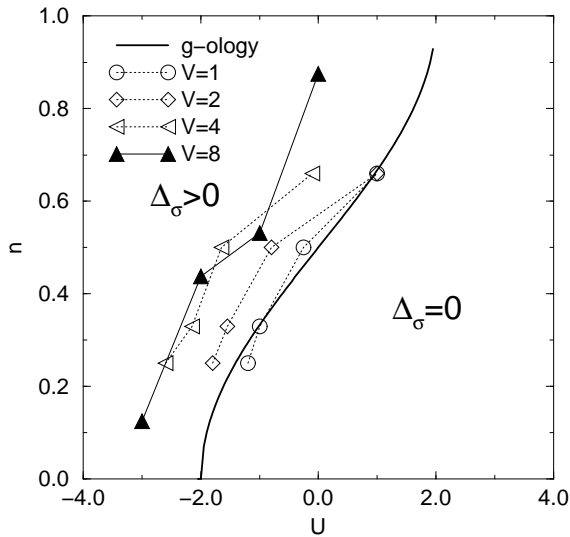}}
\caption{Summary of spin gap boundary. The solid curve 
is the g-ology result for $V=1$.
Hollow symbols connected by dotted lines are results of ED
calculations and finite-size scaling for the values of $V$ indicated. 
Solid symbols connected by lines are QMC results for 32 or 64 sites.}
\label{spin_gap_summary}
\end{figure}

When interpreting the complete
ground state phase diagram, it is useful to consider the development 
of the spin gap boundary as $V$  increases from $V=0$. For the case 
$V=0$, there is a spin gap (and dominate singlet superconducting
fluctuations) for all negative 
$U$. For weak coupling, g-ology results may be used, and they predict a spin
gap for $g_1=U+2V \cos(\pi n) < 0$. For larger $U$ and $V$ we have
used both QMC and Lancz\"os ED. In the ED calculations, the gap was 
computed from the ground state energies of the system, and the system
with one spin flipped: $\Delta_s=E(N,n_\uparrow,n_\downarrow)-E(N,n_\uparrow+1,
n_\downarrow-1)$. For each filling, two or three system sizes and finite-size
scaling of the gap values versus $1/N$ were used. The QMC estimate for the
spin gap boundary was extracted from structure factor results such as 
those shown in Fig.~\ref{spin_gap}. We consider a system to be gapped
in cases where the $S_\sigma (q)$ curve falls below the line with
slope $\pi$ as $q \to 0$. The QMC and ED results are combined in
Figure \ref{spin_gap_summary}. One can see that (i) for weak
interactions the ED results are quite close to the g--ology predictions;
and (ii) importantly, as $V$ increases, the spin-gap boundary line moves
towards the infinite $V$ result of $U=-4$. 

\begin{figure}
\centerline{\epsfig{width=3.2in,file=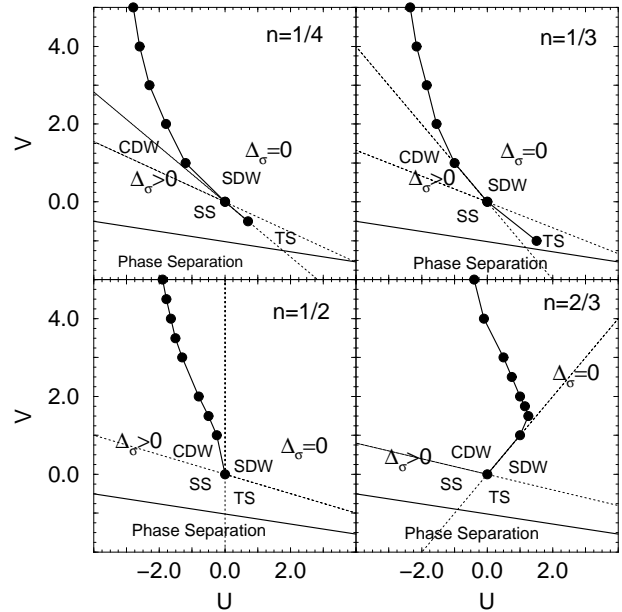}}
\caption{The spin gap region as a function of $U$ and $V$ for
$n=1/4$, $n=1/3$, $n=1/2$, and $n=2/3$. The dotted lines indicate the
phase boundaries from g-ology, and the solid circles are the results
of our exact diagonalizations. The spin gap is present to the left
of the boundaries shown. The approximative phase separation boundaries for 
negative $U$ calculated in Refs~\cite{big2,big} are also shown.}
\label{spin_gap_ED}
\end{figure}

For purposes of comparison to previous results on the spin gap in the
parameter region $V > 0$,\cite{pm,so} we present ED results graphed in
the $(U,V)$ plane for various fillings in Figure \ref{spin_gap_ED}.
In this figure we also show the g-ology predictions for the dominant
fluctuations, as well as the phase separation boundary for the $V < 0$ regime, 
which was calculated previously. \cite{big2} Our spin gap boundary 
for $n=1/2$ agrees closely with previous numerical work presented
by PM. \cite{pm} However, our boundary for $n=2/3$ is significantly 
different from that obtained previously, \cite{so}
with our results placing the 
spin gap boundary further towards the negative--U side of the phase diagram. 

\section{Discussion and Conclusion}

In the preceding sections,
we have explored numerically the $V \gg |U|\sim 1$ parameter
region of the 1D extended Hubbard model for
a wide range of band fillings. Our numerical approaches included
three different forms of QMC simulations on systems
of up to 128 sites and Lancz\"os exact diagonalizations
and finite-size scaling for systems of up to 16 sites.
In addition, we relied on some analytic
results for $V\rightarrow \infty$.
\begin{figure}
\centerline{\epsfig{width=3.0in,file=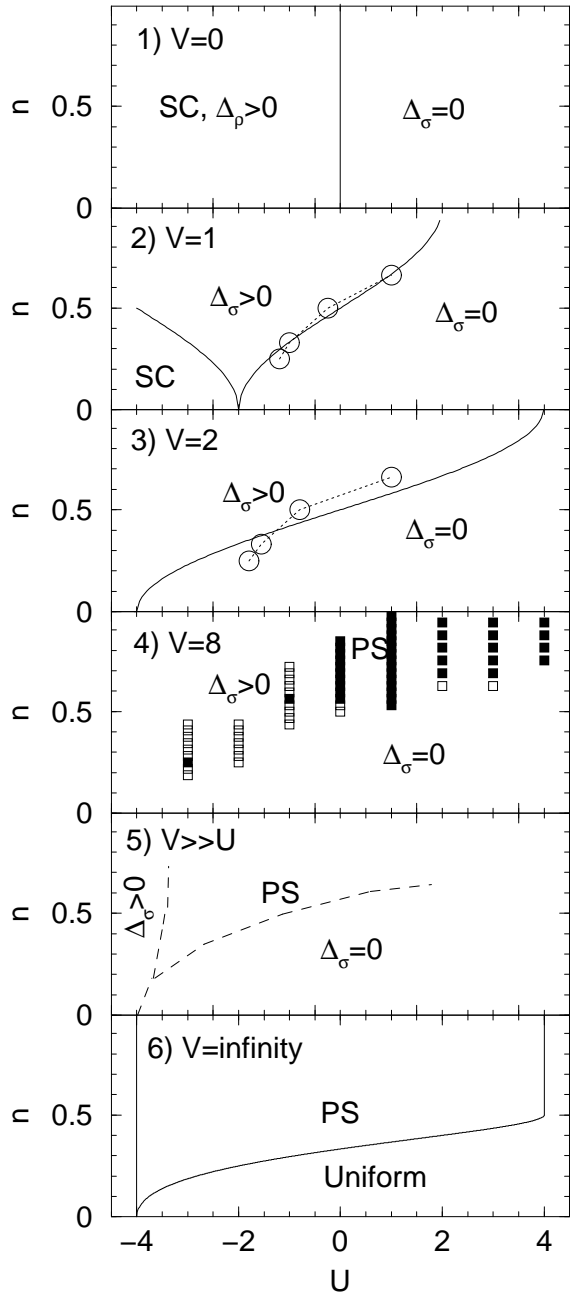}}
\caption{Phase diagram of EHM in the region $-4<U<4$. Circles denote
exact diagonalization data. The curves for $V=0$, $V=1$, and $V=2$ are from 
g-ology. At $V=8$, filled boxes are phase separation
points we have confirmed [via a
Maxwell construction or the $q\to 0$ behavior of $S_\rho(q)$ using
QMC]. Open boxes are also most likely correspond to phase separation,
but our results here are less certain. The full phase-separated region is 
inside a curve enclosing all these points. The curves shown in panel 5 
are only schematic. The curves in panel 6 are the exact solution 
for $V\rightarrow \infty$}.
\label{phase_diagram}
\end{figure}%

A complete quantitative mapping of the phase diagram even
for the restricted region $V \gg |U|\sim 1$ would require enormous
numerical work to determine precisely all phase boundaries for many
values of the parameters. Although we have not carried out such a program in 
the present study, we believe that we have developed an accurate qualitative
(with quantitative results for specific parameters) picture of the
phase diagram in a parameter 
region of the EHM about which previously little has been understood. The 
considerable complexity we have found in the model serves to illustrate 
the great care that must be taken in interpreting numerical data, in particular
those used as signals of superconductivity and phase separation
in exact diagonalization results for small lattices.

Our most important conclusion is that phase separation extends to
much lower values of $V$ than previously reported. As a result, the
Luttinger liquid exponent $K_\rho$ does not exceed one before
phase separation, and hence the ground state is not dominated by
superconducting fluctuations. This resolves the difficulties in
interpretations of previous exact diagonalization results,\cite{pm}
which indicated an extended region of gapless spin excitations and 
$K_\rho > 1$, which taken together would correspond to dominant
triplet pairing correlations. Since the naive picture of superconductivity 
in the $V \gg |U|$ region involves singlet on-site pairs, one would rather
have expected a spin gap to accompany $K_\rho > 1$. Our results
explain this puzzling result as simply due to difficulties in
detecting phase separation in small systems. It should be noted that
our results for $K_\rho$ in the uniform phase agree quite well with
the previous estimates \cite{mz,pm} up to our calculated phase
separation boundary.

To summarize our results for the EHM phase diagram for the $V\gg |U|$
region, it is most useful to examine how the features (phase
separation, spin gap, possible superconductivity,
etc., {\it evolve} together from the more familiar
$V \sim 0$ region toward $V=\infty$. Further, since
exact results exist in both the $V=0$ and $V=\infty$ limits,
following the evolution of the various boundaries
among the phases from the two known solutions is very helpful in understanding
the global evolution of the phase diagram.
Figure \ref{phase_diagram} combines our numerical data,
the exact limiting cases, and qualitative considerations to
summarize our results. The numbered comments below correspond
to the panels of figure \ref{phase_diagram}, labeled from top (1)
to bottom (6):

\begin{enumerate}

\item ($V=0$) The EHM here reduces to the standard Hubbard model
        and is spin gapped with dominant superconducting
        fluctuations for all $U<0$. 

\item ($V=1$) As $V$ increases from zero, the single (vertical) boundary
        that for V=0 separates the superconducting/spin-gapped phase from the 
        Luttinger liquid/non-spin-gapped splits into two separate boundaries,
        producing a central spin-gapped but non-superconducting
        region, which extends into both positive and negative $U$. 
        A region of singlet superconductivity
        is also present. Results of g-ology
        agree well with exact diagonalization at this weak coupling.

\item ($V=2$) For this intermediate value of $V$, the ``standard''
        superconducting region has disappeared from our parameter
        region, moving to larger
        negative $U$ values. The spin gap region also recedes further towards
        negative $U$.

\item ($V=8$) At large $V$, a region of phase separation has entered.
        Based on the
        movement of the phase-separated region towards negative $U$ as $V$
        increases, phase separation
        probably first enters near the end of the spin gap line, i.e.
        near $n=1$ and between $U=0$ and $U=4$. For $U<0$, the spin
        gap boundary remains largely unchanged from weak coupling. At positive
        $U$, the high-density phase consists of pairs on every other site.
        For negative $U$, the high-density phase expands slightly,
        as the dominant wavevector is no longer at $q=\pi$. In
        addition, the region of phase separation
        exists over a limited region of $n$ and
        vanishes near half filling, with the homogeneous phase re-entering.
        For small fillings, $4k_F$ charge correlations dominate,
        while near half filling $2k_F$ charge correlations dominate.
        At intermediate fillings we find an anomalous, non-Luttinger
        liquid peak that
        is non-divergent with increasing system size or decreasing
        temperature. This peak appears to reflect CDW fluctuations
        involving long-lived pairs that repel each other as well as the
        unpaired particles. Our data, which clearly show phase
        separation at this
        $V$, show no convincing signs of dominant superconducting
        correlations in the uniform phase, leading us to conclude 
        that any region of superconductivity for $V \gg |U|$ is very
        small indeed, or, more likely, does not exist.

\item ($V\gg U$) At still larger $V$, we expect the region
        of phase separation to grow towards the point $U=-4$, $n=0$.
        We show a schematic phase diagram in this case.

\item ($V=\infty$) The point connecting the spin gap and phase
        separation boundaries
        has moved to $(U,n) = (-4,0)$. The spin gap boundary is now
        the vertical
        line from $n=0$ to $n=1$ at $U=-4$. The phase separation
        region consists of either effective
        spinless fermions or pairs, and the high density phase
        has pairs on every other site.

\end{enumerate}

One may argue that our numerical data cannot exclude the existence of 
an extremely narrow strip of superconductivity preceding phase
separation. However, in the non-spin-gapped region this seems unlikely 
in view of the obvious difficulty in accounting for dominant triplet
superconducting fluctuations in the $V \gg |U|$ region. The exact 
$V \to \infty$ solution also provides a counter-argument: below the phase
separation boundary ($n < n_{\rm  LD}$), 
the system is mapped onto spinless fermions, and the dominant correlations
are then clearly not superconducting in the neighborhood of the
phase separation boundary.
For large but finite $V$, we find numerically that $K_\rho$ does not
appear to exceed unity close to the phase separation boundary. We instead find
strong $2k_{\rm F}$ spin fluctuations that do not vanish as the
phase separation boundary is crossed. Hence we believe that the
system is dominated by SDW
fluctuations adjacent to the phase separation boundary in cases where there is
no spin gap. On the other hand, for $U < 0$ we find $K_\rho \approx 1$
in a narrow region for which we cannot definitely conclude that
the system is phase separated. This ambiguous region coincides with our
calculated spin gap boundary, and is essentially the region of the
open boxes in panel 4 of Fig.~\ref{phase_diagram}. Hence, if
superconducting
fluctuations indeed dominate in this region, they would be of the
singlet type, which is what one would expect. However, also in this
case it appears more likely that $K_\rho$ does not exceed unity at
the phase separation boundary. The $V \to \infty$ case again
provides support for
this result: On the left-hand side of the phase-separated region
shown in panel 6 of
Fig.~\ref{phase_diagram} (i.e., $U < -4$), the system contains only
pairs, and for large but finite $V$ exhibits strong $2k_{\rm F}$ CDW 
fluctuations. Dominant superconducting
fluctuations do appear in this region for 
small values of $V$ but are replaced by CDW fluctuations for moderate 
$V$ (see panels 2 and 3 of Fig.~\ref{phase_diagram}). It then appears
unlikely that a small region of dominant superconducting fluctuations in the 
region $U \sim (-3,-1)$ and $n \sim (0.2,0.5)$ would re-appear as $V$
is increased and then again vanish as $V \to \infty$, but we
do not have data that can definitely exclude this scenario.

Although we believe our results establish that phase separation,
rather than superconductivity, dominates the $V\gg |U|$ region
of the EHM, there remain a number of interesting open
questions about some of the ``exotic'' non-Luttinger liquid effects
we have observed. In particular, two such questions
involve (i) the exact nature of the state
corresponding to the anomalous $4k^{eff}_F$ peak in the
charge susceptibility and (ii) a more detailed understanding of the
``re-entrant'' phase separation behavior compared to the normal
$U >0$ phase separation, where the high-density phase is
at (the naively expected) half-filled density. We have presented
qualitative interpretations
of these ``exotic'' effects, but they may warrant further study,
focusing on when they can appear, what simple 
effective model (if any) can be used to describe them, the relation 
between phase separation and the non-Luttinger liquid peak, and how
this peak evolves with system size and temperature. One possible method
to access this parameter region would be to study
numerically an effective model including paired and single electrons 
similar to PM's $V\rightarrow \infty$ effective model. Such an effective
model would eliminate electrons on nearest-neighbor sites, which
could be additionally taken into account perturbatively,
if necessary. We are currently exploring this approach.

We are grateful to Steve Hellberg, Steve Kivelson, and
Reinhard Noack for useful
discussions. This work is supported by the NSF under grant DMR-97-12765. RTC 
acknowledges support of a NSF GRT fellowship. The numerical calculations 
were performed in part at the NCSA.

\end{document}